\documentclass[a4paper,11pt]{article}

\usepackage{multicol}
\usepackage[latin1]{inputenc}
\usepackage{authblk}
\usepackage{amsmath}
\usepackage{amsfonts}
\usepackage{amssymb}
\usepackage{graphicx}
\usepackage{appendix}
\usepackage{caption}
\usepackage{subfigure}
\usepackage{float}
\usepackage{enumerate}
\usepackage{latexsym}
\usepackage{epsfig}
\usepackage[usenames]{color}
\usepackage[colorlinks=true]{hyperref}
\usepackage{pdfpages}
\usepackage{setspace}
\singlespace
\newcommand{\ssc}{\mathcal{S}}
\newcommand{\st}{\mathcal{T}}

\graphicspath{ {figures/} }

\title{Hemispherical asymmetry from an isotropy violating stochastic gravitational wave background} 
\author{Suvodip Mukherjee\footnote{suvodip@iucaa.ernet.in}\\
IUCAA, Post Bag 4, Ganeshkhind, Pune-411007, India\\}

\date{\today}
\begin{document}

\maketitle{} 

\pagenumbering{arabic}
\thispagestyle{plain}
\markright{}

\begin{abstract}

Measurement of Cosmic Microwave Background (CMB) temperature by Planck has resulted in extremely tight constraints on the $\Lambda$CDM model. However the data indicates a evidence of dipole modulated temperature fluctuations at large angular scale which is beyond the standard Statistically Isotropic (SI) $\Lambda$CDM model. The signal measured by Planck requires a \textit{scale dependent} modulation amplitude that is beyond the scope of the phenomenological model considered by Planck. We propose a phenomenological model with mixed modulation field for scalar and tensor perturbations which affect the temperature fluctuations at large angular scales. Hence this model is a possible route to explain the \textit {scale dependent} nature of the modulation field. The salient prediction of this model is the direction dependent tensor to scalar ratio which results in anisotropic Stochastic Gravitational Wave Background (SGWB). This feature is potentially measurable from the $B$-mode polarization map of Planck and BICEP-2 and leads to determination of the modulation strength. Measurability of SI violated polarization field due to this model is estimated for Planck and PRISM. Absence of the signal in the polarization field can restrict the viability of the model.

\end{abstract}

\begin{multicols}{2}
\section{Introduction}
The sustained improvement in sensitivity of the instrumental noise over the last couple of decades has made Cosmic Microwave Background (CMB) a very powerful probe of our Universe. Measurement of temperature power spectra from Planck \cite{Planck4, Planck_param} is explained with the minimal $\Lambda$CDM model  at angular scales smaller than $2\,^{\circ}$ (multipole $l>50$). However, Planck \cite{planck23} and also WMAP \cite{erikson, hoftuft} detected significantly more fluctuations in the CMB temperature field in one hemisphere centered in the direction $\hat p\, = \,(l = 217.5^\circ, b= -20.2^\circ)$. This feature cannot be accommodated within the standard Statistically Isotropic (SI) cosmological models. 

This observed SI violation is phenomenologically modelled by the dipole modulation of the total temperature field and it yields a scale dependent modulation present only at  large angular scales $(\theta \geq 3^\circ)$ \cite{planck23}, which is beyond the scope of dipole modulation model. Hence the observed scale dependent feature of the dipole asymmetry indicates that this model is not the true model to explain the observed SI violation. Also detection of similar anisotropic feature from two independent experiments (WMAP and Planck) suggests a cosmological origin of the modulation field in favour of ignored systematic effect \cite{hansen, eriksen1, akrami, miguel}.

To investigate the cosmological origin of this dipole modulation which is present only at  large angular scale,  we propose a phenomenological model in which the SI violated features are prominent in both scalar and tensor perturbation. The modulated tensor perturbations leads to anisotropic Stochastic Gravitational Wave Background (SGWB) which has negligible contribution at small angular scales  
 and hence can naturally lead to the observed scale dependence. This model affects both the temperature and polarization field of CMB and leads to a direction dependent tensor to scalar ratio $r(\hat n)$ which is different from the all sky average value of tensor to scalar ratio $\langle r \rangle$.

From the measurement of $B$-mode polarization from small-sky experiment BICEP-2 \cite{bicep2} and full sky experiment Planck, the modulation strength for the tensor part can be estimated. 
\section{BipoSH representation for temperature and polarization}\label{2}
The $\Lambda$CDM model of cosmology assumes the universe to be statistically isotropic and hence predicts the harmonic space covariance matrix to be diagonal $\langle X_{l_1 m_1} X^{'*}_{l_2 m_2} \rangle = C^{XX'}_{l_1} \delta_{l_1 l_2}\delta_{m_1 m_2}$. However this assumption of isotropy is now being rigorously tested with almost full sky measurements of CMB temperature anisotropies. The basic idea behind this study is to search for statistically significant power in off-diagonal elements of the covariance matrix $\langle X_{l_1m_1} X'^*_{l_2m_2} \rangle$.  The Bipolar Spherical Harmonic (BipoSH) coefficients introduced in this field by Hajian and Souradeep \cite{ts, tsa, tsb} form a very convenient basis to study deviations from statistical isotropy. In the absence of the isotropy assumption the two point correlation of the CMB temperature (T) and polarization fields ($E$ \& $B$) can be most generally expressed in terms of these basis functions. The two point correlation of CMB anisotropy $C^{XX'}(\hat n_1, \hat n_2) = \langle X(\hat n_1) X'(\hat n_2)\rangle$  can be expressed most generally as
\begin{align}\label{eqbi2a}
\begin{split}
C^{XX'}(\hat n_1, \hat n_2)=&\\ \sum_{JN l_1 l_2} & A^{JN}_{l_1l_2|XX'}\{Y_{l_{1}}(\hat n_{1})\otimes Y_{l_{2}}(\hat n_{2})\}_{JN},
\end{split}
\end{align}
where $A^{JN}_{l_1l_2|XX'}$ are the BipoSH coefficients for $X= \,  T, E, B$ and can be related to the 
harmonic space covariance matrix through the following expression,
\begin{align}\label{eqbi2}
\big\langle X_{l_1m_1} X'^*_{l_2m_2}\big\rangle = \sum_{JN} A^{JN}_{l_1l_2|XX'}(-1)^{-m_2}C^{JN}_{l_1 m_1 l_2 -m_2} ,
\end{align}
%
where, $C^{JN}_{l_1m_1l_2m_2}$ are the Clebsch-Gordan coefficients.  From a given temperature or polarization map, BipoSH coefficients can be obtained by inverting the above relation 
\begin{align}\label{eqbi2}
A^{JN}_{l_1l_2|XX'} = \sum_{m_1 m_2} (-1)^{-m_2}\big\langle X_{l_1m_1} X'^*_{l_2m_2}\big\rangle C^{JN}_{l_1 m_1 l_2 -m_2} ,
\end{align}
In the isotropic case, the only non-vanishing BipoSH coefficients are given by the expression : $A^{JN}_{l_1 l_2 |XX'} = (-1)^{l_1} \sqrt{(2l_1+1)}C^{XX'}_{l_1} \delta_{l_1 l_2} \delta_{J0}\delta_{N0} $. Measurement of any non-zero value of BipoSH coefficients for $J \neq 0$ is the signature of violation of statistical isotropy.

\section{Mixed modulation in scalar and tensor perturbation}\label{3}
The measurements of the CMB temperature anisotropies by Planck indicates a  dipole modulation of the CMB sky on large angular scale with a Signal to Noise Ratio (SNR) of $3.7$ \cite{planck23}. However it is known that the dipole modulation of the CMB sky cannot be the true model as it requires scale dependent dipole modulation amplitude to explain the observations \cite{planck23}. 
\begin{figure}[H]
\centering
\includegraphics[width=3.0in,keepaspectratio=true]{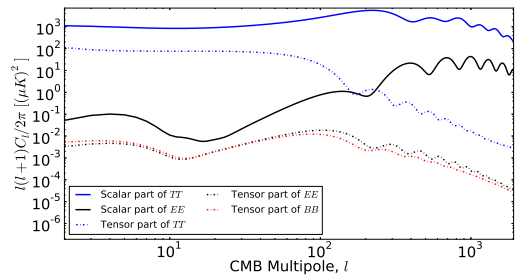}
\caption{The figure shows the scalar and tensor metric perturbation contributions to the CMB angular power spectra generated using CAMB \cite{camb} with the value of tensor to scalar ratio $(r)$ as $0.2$. Note that while the $C_l^{TT}$ and $C_l^{EE}$ have large contributions coming from scalar metric perturbations, the $C_l^{BB}$ spectrum on large angular scales is purely sourced by tensors. }\label{scalar-tensor-contribution}
\end{figure}
To explain this we consider a model with different amount of modulation in scalar and tensor part of the metric fluctuation and we refer this as \textit{Mixed Modulation} (MM). Since, the contribution to temperature anisotropy arising from the tensor part of the metric fluctuations decay at small angular scale as shown in Fig.~\ref{scalar-tensor-contribution}, it is expected for the tensor modulation field also to be insignificant at such scales. As a result this model can naturally mimic the observed scale dependent dipole modulation. Recently, different sort of effects from tensor perturbations to account for the observed anomalies is also showed by several authors \cite{chen, zibin, emami}. A variety of models like direction dependent scalar spectral index ($n_s$), optical depth ($\tau$), gravitational waves and isocurvarture perturbations,  which can lead to a dipole asymmetry is  discussed by Dai et al. \cite{dai}. 
Effect of the MM  on temperature and polarization field can be written in terms of a mixing factor $\alpha$. On assuming that $\alpha$ part of the observed dipole modulation is from tensor perturbation and the remaining $1-\alpha$ part is from scalar perturbation, the total effect on temperature and polarization field can be written as, 
\begin{multline}\label{modulate-mixed}
\tilde{X}(\hat n)= [1+ (1-\alpha)M_{\ssc}(\hat n)]X^{\ssc}(\hat n) +\\ [1+ \alpha M_{\st}(\hat n)] X^{\st}(\hat n) \,, 
\end{multline}
where, $X \in {T,E,B}$. The superscripts $\ssc$ \& $\st$ denote the contribution to CMB fields arising from the scalar and tensor components of the metric perturbations respectively. $M_{\ssc}$ ($M_{\st}$) denotes the modulation field for scalar (tensor) part which can be determined from the Planck temperature field by setting $\alpha$ either to zero or one. Our model becomes equivalent to the model considered by Planck \cite{planck23} by choosing the value of $\alpha= M_{\ssc}/(M_{\ssc}+M_{\st})$ which makes the values for $(1-\alpha)M_{\ssc}$ and $\alpha M_{\st}$ equal to $M_{\ssc}M_{\st}/(M_{\ssc}+M_{\st})$. Hence the MM  model considered by us is a more general model compared to Planck, which has considered only a special case of it. 
For $\alpha > 0.2$, the signal strength $(1-\alpha)M_{\ssc}$ does not remains significant (SNR $<3$ ) at small angular scales and hence leads to a scale dependent modulation field. As a result this model has potential to explain the complete features of the observed non-SI signal. 
This model with modulation in the tensor part is equivalent to the SI violated Stochastic Gravitational Wave Background (SGWB), which results in direction dependent tensor to scalar ratio $(r)$. A spatially varying tensor to scalar ratio $r$ is recently discussed by Chluba et al. \cite{chluba}. For a small patch of the sky along the direction $\hat n$ we use the technique discussed by Hivon et al. \cite{master} to obtain the direction dependent $r(\hat n)$ as,
\begin{equation} \label{modulate-r1}
r(\hat n)=r_o \left[1+ \alpha M_{\st}(\hat{n}) \right]^2 \,,
\end{equation}
where $r_0$ is the unmodulated tensor to scalar ratio. Then the all sky averaged value of tensor to scalar ratio (denoted by $\langle r(\hat n) \rangle$) becomes,
\begin{equation} \label{modulate-avg-r1}
\langle r(\hat n) \rangle=r_o \langle (1+ \alpha M_{\st}(\hat{n}))^2\rangle \,,
\end{equation} 
Since the salient feature of MM  model is the SI violated tensor field which leads to dominant non-SI signal at large angular scales, we estimate the tensor modulation field $M_{\st} (\hat n)$ from Planck temperature map in the next section by setting $\alpha=1$ in Eq.~\ref{modulate-mixed}.
 \section{Determination of Tensor modulation field from Planck and BICEP-2}\label{4a}
\subsection{Estimating $M_{\st}$ from the observed Planck temperature map}\label{4}
Here we investigate the possibility of statistically anisotropic signal by modulating only the temperature anisotropies arising from tensor part of the metric fluctuations by choosing $\alpha=1$. Then Eq.~\ref{modulate-mixed} becomes 
\begin{eqnarray}  \label{modulate-cmb}
\tilde{T}(\hat n)= T^{\ssc}(\hat n) + [1+ M_{\st}(\hat n)]T^{\st}(\hat n) \,.
\end{eqnarray}

Using the BipoSH formalism mentioned in Sec. \ref{2}, we determine the nature of $M_{\st}(\hat n)$ from the observed temperature field by Planck.  
The BipoSH coefficients for temperature anisotropy is 
\begin{equation}\label{bips}
A^{JN}_{ll'} =  M_{JN} \,^{\st}S^{J}_{ll'}  ~~~~ \forall ~~J \neq 0 \,,
\end{equation}
%
where, $M_{JN}$ are the spherical harmonic coefficients of the modulation field $M_{\st}$ and $^{\st}S^{J}_{ll'}$ is the shape factor given by
\begin{equation} \label{shape-functions}
^{\st}S^{J}_{ll'|TT} = \frac{\Pi_{ll'}}{\Pi_J} \frac{\left[ \,^{\st}C^{TT}_{l}+\, ^{\st}C^{TT}_{l'} \right]}{\sqrt{4\pi}} \mathcal{C}^{J0}_{l0l'\,0} \,, 
\end{equation}
where $\Pi_{ll'} = \sqrt{(2l+1)(2l'+1)}$ and $^{\st}C^{TT}_{l}$ denotes the angular power spectrum of the tensor modes of CMB temperature field. Since, the angular power spectra depends upon the value of tensor to scalar ratio $r_0$, the BipoSH spectra also strongly depends upon the value of $r_0$. So, precise estimation of $r_0$ is essential for correctly estimating the modulation strength from the BipoSH spectra.  

Finally we derive a minimum variance estimator for reconstructing the modulation field following an identical procedure as described in \cite{planck23, doppler, hu2, duncan2}. The minimum variance estimator derived is given by the following expression,
\begin{equation}\label{mvestimator}
\hat{M}_{JN} = \frac{\sum_{ll'} S^{J}_{ll'} A^{JN}_{ll'} / \sigma^2_{A^{JN}_{ll'}} }{\sum_{ll'}(S^{J}_{ll'})^2 / \sigma^2_{A^{JN}_{ll'}}} \,,
\end{equation}
where variance of the BipoSH coefficients $\sigma^2_{A^{JN}_{ll'|XX'}} = \langle A^{*JN}_{ll'}A^{JN}_{ll'}\rangle$ can be expressed in terms of the total CMB angular power spectrum for an SI sky \cite{doppler}.

\begin{figure}[H]
\centering
\includegraphics[width=3.6in,keepaspectratio=true]{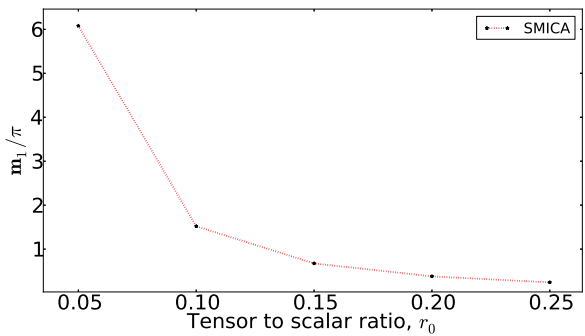}
\caption{
Estimation of tensor modulation power $m_1/\pi$ for different values of $r_0$ from Planck SMICA map. This shows a decrease in $m_1$ with increase in $r_0$. 
}\label{fig:rm}
\end{figure}

The significance of the detections of the modulation field harmonics is evaluated by estimating the reconstruction noise properties of the minimum variance estimator. 
It can be shown that the reconstruction noise in each harmonic mode $J$ of the modulation field is given by the following expression, 
\begin{equation}\label{eqnc8}
N_J= \bigg[\sum_{ll'}\frac{(S^{J}_{ll'})^2}{\sigma^2_{A^{JN}_{ll'}}}\bigg]^{-1} \,.
\end{equation}
To mimic the observed dipole asymmetry, the amplitude of the modulation strength for tensor field can be estimated from the Planck SMICA map \cite{Planck_map}. This amplitude depends upon the value of unmodulated tensor to scalar ratio $r_0$. Since, the value of $r_0$ is not measured precisely, we estimate the $m_1/\pi= \sum _{N} \frac{|M_{1N}|^2}{3\pi}$ for different $r_0$ as shown in Fig. \ref{fig:rm}.

\begin{figure*}[t]
\subfigure[]{
\includegraphics[width=3.2in,keepaspectratio=true]{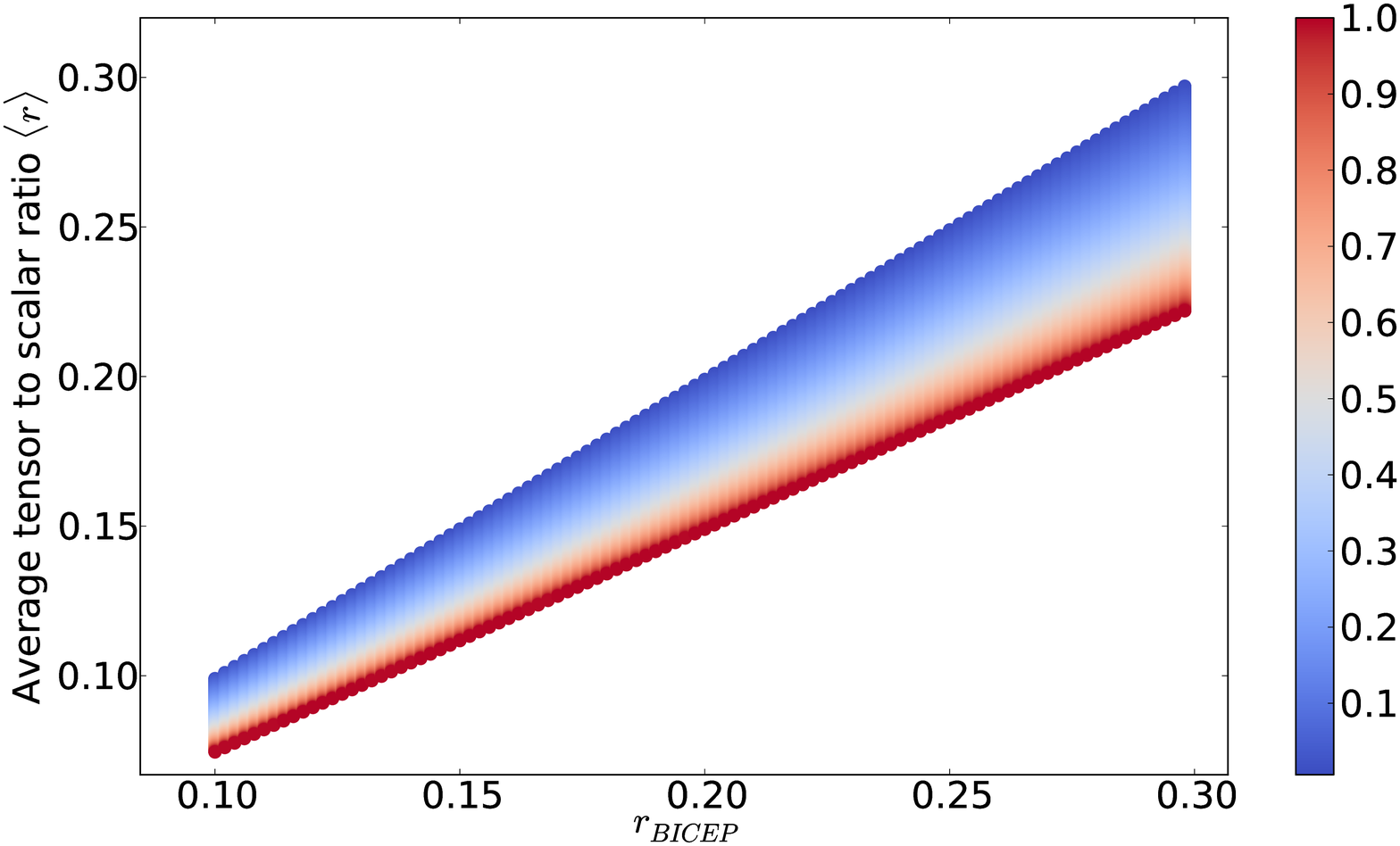}\label{fig:rb-ravg-alpha}
}
\subfigure[]{
\includegraphics[width=3.2in,keepaspectratio=true]{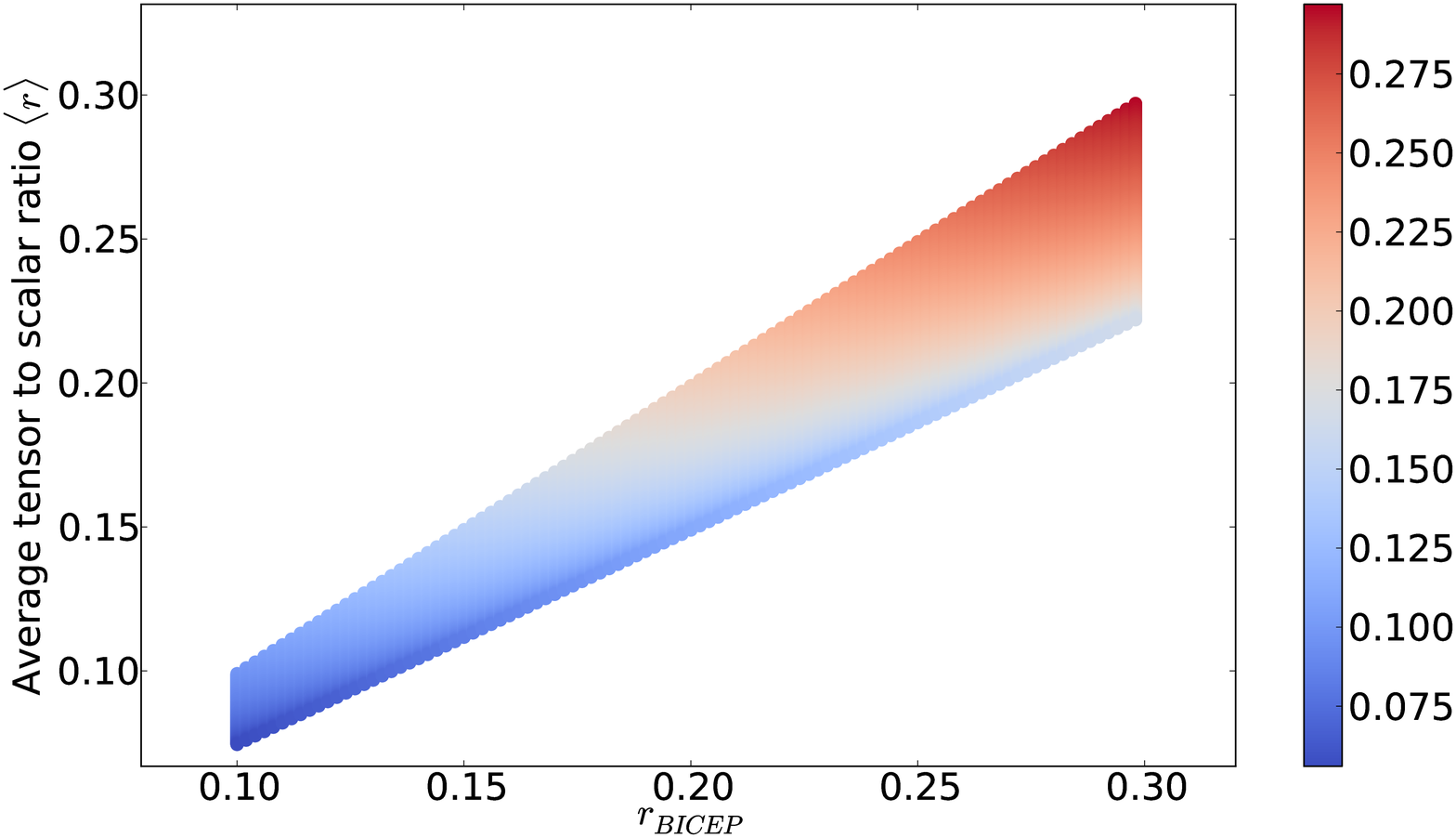}\label{fig:rb-ravg-r0}
}
\caption{(a) In this plot we show the possible range of  $r_{\rm BICEP}, \langle r \rangle$ for different values of modulation strength  $\alpha A$ (shown in colour bar). This shows that with the measured value of $r_{\rm BICEP}$ in the BICEP patch and all sky average value $\langle r \rangle$, we can estimate the value of $\alpha$ uniquely. (b) In this plot we show the determination of $r_0$ (shown in colour bar) from the measurement of $r_{\rm BICEP}$ and  $\langle r \rangle$. From the measurement of a particular value of $r_{\rm BICEP}$ and  $\langle r \rangle$, both the values of $\alpha A$ and $r_0$ can be determined.}
\end{figure*}
 Since the only significant power is seen in the dipolar $(J=1)$ mode of the modulation field, we can take the modulation field of the form, $M_{\st}(\hat{n}) = A \,\hat{p}\cdot\hat{n}$, where A denotes the amplitude of the dipole and $\hat{p}$ is the dipole direction. It can be shown that the dipole amplitude $A$ is related to the power in $J=1$ mode by $A= 1.5\sqrt{{m}_1/\pi}$. By estimating the value of $M_{\st}$, we can determine the mixing factor $\alpha$ from polarization spectra.
\subsection{Measuring the mixing factor $\alpha$ from polarization map of Planck and BICEP-2}\label{5}
Tensor part of the mixed modulation leads to observable effect in the $B$ mode polarization of CMB. In the presence of non-zero $\alpha$ with the dipolar nature of tensor modulation shown in Sec. \ref{4}, the tensor to scalar ratio in a small patch of the sky shown in Eq. \ref{modulate-r1} becomes,
\begin{equation} \label{modulate-mod-r}
r(\hat n)=r_o \left[1+ \alpha A\, \hat p. \hat n \right]^2 \,.
\end{equation}

The all sky average $r$ for a dipolar modulation field then becomes,
\begin{equation} \label{modulate-avg-r}
\langle r(\hat n) \rangle=r_o (1+ (\alpha A)^2/3) \,,
\end{equation}
Using these two equations, we can estimate the value of $\alpha$ and $r_0$ from the CMB missions like BICEP-2 and Planck. 

BICEP-2 \cite{bicep2} has measured the polarized CMB sky, around the direction $\hat n_0\, = (l= 314^\circ, b= -59^\circ)$ and has claimed a detection of  $r_{\rm BICEP}=0.2^{+0.07}_{-0.05}$. However, the value of the $r$ in the BICEP field may alter on considering  foregrounds from the dust in the mentioned patch which was recently pointed out by Planck \cite{Planck_dust}. 

Considering the angular separation between BICEP-2 patch and the dipole direction $(\hat p)$ as $\theta = \cos^{-1}( \hat p.\hat n_0) \approx 75^\circ$, we show the allowed possible values for $r_{\rm BICEP}$ and $\langle r \rangle$ for different $\alpha A$ and $r_0$ in Fig.~\ref{fig:rb-ravg-alpha} and \ref{fig:rb-ravg-r0} respectively. This plot indicates that a stronger (weaker) modulation strength $\alpha A$ can lead to a larger (smaller) discrepancy between the measurement of $r_{\rm BICEP}$ and $\langle r \rangle$.   A precise measurement of tensor to scalar ratio from BICEP-2 and Planck can determine $\alpha A$ and $r_0$. If the value of $\alpha$ is consistent with zero then the MM  model can be ruled out. 
\section{Forecast for measuring SI violated $B$ mode polarization from Planck}\label{6}
The MM  model with a significant contribution from tensor perturbation  can lead to the scale dependent dipole modulation in the temperature field. This model can also result in SI violation in the polarization field of CMB and hence non-zero value of BipoSH spectra for polarization. 
Since, the main feature of MM  model is the presence of SI violated tensor modulation, we make a forecast of measuring the BipoSH spectra for polarization from Planck arising from tensor part of this model. Detection of this modulation signal from $BB$ BipoSH spectra strengthens the viability of MM  model. 

\begin{figure}[H]
\centering
\includegraphics[width=3.2in,keepaspectratio=true]{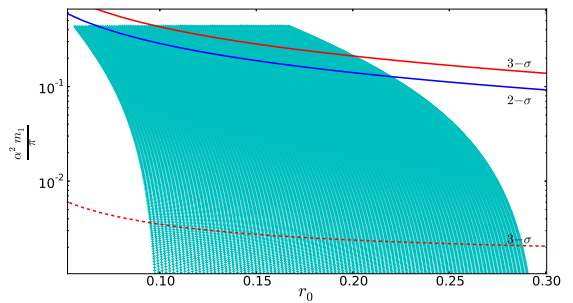}
\caption{The dipole modulation $\frac{\alpha^2m_1}{\pi}$ is plotted for different values of $r_0$. In blue and red we plot the $2\sigma$ and $3\sigma$ curve of the reconstruction noise $(\sigma = N_1/\pi)$ upto $l_{\rm max} = 64$ as a function of $r_0$ after incorporating Planck instrumental noise for frequency channel $\nu=143 \,\rm GHz$. 
In the dashed red line we depict $3\sigma$ curve of the reconstruction noise for PRISM. The effect of this model is readily measurable in the $B$-mode BipoSH spectra from PRISM for a wide range of $\frac{\alpha^2m_1}{\pi}$.}\label{fig:4}
\end{figure}

The measurability of BipoSH spectra can be calculated using the reconstruction noise $N_1$ mentioned in Eq. \ref{eqnc8} for the dipole modulated $E$ and $B$ modes. The reconstruction noise $N_1$ for $EE$ and $TE$ spectrum are much larger than the reconstruction noise for $BB$ spectrum. This is due to the large contribution from the scalar perturbations to the cosmic variance for $EE$ and $TE$, which is absent in the $B$ mode polarization. 
Hence we estimate the detectability of the effective modulation signal $\alpha A$ only for the $B$ mode polarization. As there is no precise detection of $r_0$, we  estimate reconstruction noise $N_1$ (upto $l_{max}= 64$) for the $B$ mode with different values of $r_0$, for one of the best frequency channel $\nu = 143\, \rm GHz$ of Planck \cite{planckblue}. The $2\sigma$ and $3\sigma$ curves are plotted in blue and red solid lines respectively in Fig. \ref{fig:4} along with a range of values for $\alpha^2 m_1/\pi$ and $r_0$. With the measurement of $\alpha^2 m_1/\pi$ and corresponding $r_0$ from Planck and BICEP-2 (as mentioned in Sec. \ref{5}), we can determine the detectability of $\alpha^2 m_1/\pi$ from $B$ mode BipoSH spectrum. 
In Fig. \ref{fig:4}, the region above the $3\sigma$ curve (solid red line) is detectable with SNR $\geq 3$ from Planck. This implies that the non-SI $B$ mode polarization is detectable only for a limited range of modulation strength from Planck. However the reconstruction noise for the frequency channel $\nu= 220\, \rm GHz$ of future mission such as PRISM \cite{prism} shown by the red dashed curve in Fig.~\ref{fig:4},  is considerably  lower than Planck. As a result, PRISM can measure $B$-mode BipoSH spectra even for weak power of effective modulation. Absence of such signal from PRISM can falsify the proposed MM  model.

\section{Conclusions}\label{7}
Planck \cite{planck23} has measured a signature of Statistical Isotropy (SI) violation at large angular scales which is beyond the standard $\Lambda$CDM model. This SI violated signal is modelled as a dipole modulation of the total temperature field with a peak amplitude of $0.07$ \cite{planck23}. But this model does not capture the features correctly, since it is unable to recover  a scale dependent modulation amplitude \cite{planck23}.
 
In this paper, we propose Mixed Modulation (MM) model, having different modulation strengths for scalar and tensor perturbations that can capture the observed SI violated signal without scale dependent modulation amplitude. The different amplitudes of modulation strength for scalar and tensor can be modelled by a mixing factor $\alpha$ as shown in Eq. \ref{modulate-mixed}. Since, CMB temperature anisotropies arising from tensor perturbation decays at small angular scales (shown in Fig. \ref{scalar-tensor-contribution}), any modulation in the tensor anisotropy is also expected to decay at such scales. Hence the only contribution from Eq. \ref{modulate-mixed} at small angular scales comes from scalar perturbation part, i.e. $(1-\alpha)M_{\ssc}$ for MM  model. 

Since, the salient feature of the MM  model is the modulated tensor perturbation, it is essential to estimate the tensor part of the modulation field $M_{\st}$ from the observed temperature field by Planck \cite{planck23}. For this we consider the MM model with $\alpha=1$ (i.e. only tensor modulation) and calculate the BipoSH spectra and the corresponding minimum variance estimator for temperature field in Eq. \ref{bips} and Eq. \ref{mvestimator} respectively. The power of the amplitude of the modulation strength $m_1/\pi$ depends upon the value of tensor to scalar ratio $r_0$ used in the analysis. In Fig. \ref{fig:rm}, we plot the estimated power $m_1/\pi$ for different values of $r_0$ from observed Planck SMICA map. This indicates that for higher tensor to scalar ratio $r_0$, we need a lesser modulation strength to explain the observed amplitude of SI violation. Due to the dipolar nature of $M_{\st}$ we can express the modulation field as $M_{\st}= A \,\hat p.\hat n$, where $A$ is related to the power of the modulation field $m_1/\pi$ by $A= 1.5 \sqrt{m_1/\pi}$. This non-SI tensor perturbation implies anisotropic Stochastic Gravitational Wave Background (SGWB) which causes direction dependent tensor to scalar ratio. Different mechanisms like exotic super horizon tensor modes \cite{akbar}, dissipative processes \cite{amico}, modulated preheating \cite{bethke} can produce non-SI SGWB. Several authors \cite{namjoo, zarei} have also mentioned about other processes to produce modulated tensor perturbation. An SI violated SGWB can also arise due to an initial anisotropy during inflation \cite{mukherjee_inf}. The model with anisotropy in the inflationary dynamics considered by Mukherjee and Souradeep \cite {mukherjee_inf} leads to dominant SI violation in the tensor perturbation than in the scalar perturbation. This mechanism naturally leads to a direction dependent scalar spectral index ($n_s$) and also satisfies the results as shown by other authors \cite{dai, gorski}.
In a recent paper by Dai et al. \cite{dai}, a variety of models were discussed which can lead to the observed scale dependent dipole asymmetry. One of such model is the direction dependent $n_s$, which can lead to the observed signal. By using the MM model, the value of the anisotropy in the  scalar part reduces and as a result of which even a mild direction dependence in the value of $n_s$ can express the observed signal. Detection of $\alpha$ from the future missions can impose strong constrain on the direction dependence of $n_s$.
 
To estimate the value of mixing factor $\alpha$, we need to consider multiple independent measurements of tensor to scalar ratio $r$. On estimating the tensor to scalar ratio from BICEP-2 patch ($r_{\rm BICEP}$) and all sky average tensor to scalar ratio from Planck ($\langle r \rangle$), we can estimate the value $r_0$ and $\alpha A$ using the Eq. \ref{modulate-mod-r}, \ref{modulate-avg-r}. In Fig. \ref{fig:rb-ravg-alpha}, we plot a range of values for $r_{\rm BICEP}$ and $\langle r \rangle$ for different values of $\alpha A$. This plot indicates that for a measured value of $r_{\rm BICEP}$ and $\langle r \rangle$, $\alpha A$ can be determined. Stronger (weaker) the value of  $\alpha A$, larger (smaller) the discrepancy between $r_{\rm BICEP}$ and $\langle r \rangle$. 
Similarly for different values of $r_0$, $r_{\rm BICEP}$ and $\langle r \rangle$ is plotted in Fig. \ref{fig:rb-ravg-r0}. 
Also to investigate the signature of  tensor part of the MM  model on BipoSH spectra for polarization, we estimate the reconstruction noise $N_1$ mentioned in Eq. \ref{eqnc8} for $EE, TE$ and $BB$ spectrum with Planck instrumental noise for frequency channel $\nu=143 \,\rm GHz$. Due to large contribution from scalar perturbation in $EE$ and $TE$ spectrum, the cosmic variance for $EE$ and $TE$ are larger than $BB$ spectrum. As a result, the reconstruction noise is minimum in $BB$ and leads to a measurable signature only in the BipoSH spectra for $BB$ spectrum. Since the effective modulation power $\alpha^2 m_1/\pi$ and unmodulated tensor to scalar ratio $r_0$ are not measured precisely, in Fig. \ref{fig:4} we plot a range of values for $\alpha^2 m_1/\pi$ along with $2\sigma$ and $3\sigma$ curve for different values of $r_0$, where $\sigma$ is denoted by $N_1/\pi$. This figure shows that only a limited range of the modulation strength can have a detectable effect in Planck from $BB$ BipoSH spectrum.
However, with future missions like PRISM \cite{prism}, much lesser value of $\alpha^2m_1/\pi$ are also detectable with a high SNR. In Fig. \ref{fig:4}, we plot the value of reconstruction noise for frequency channel $\nu=220 \,\rm GHz$ of PRISM by red dashed curve.

The proposed MM  model generates SI violated tensor perturbation, which implies an SI violated Stochastic Gravitational Wave Background (SGWB). A similar direction dependent tensor to scalar ratio was also discussed earlier  by Chluba et al \cite{chluba}. Our model leads to several observable signatures in the $B$-mode polarization spectra which are measurable from experiments like BICEP-2, Planck and PRISM. A detailed study of this model is required using all future data to understand the cosmological origin of the observed SI violation.

\textbf{Acknowledgement:}
I would like to thank Professor Tarun Souradeep for useful discussions and comments. All the computational works are carried out in the High Performance Computing facility at IUCAA. I acknowledge Council of Scientific \& Industrial Research (CSIR), India for the financial support.

\end{multicols}
\end{document}